\documentclass[letterpaper,10pt]{article}
\usepackage{osameet2}
\usepackage[font=footnotesize,labelfont=bf]{caption}
\usepackage{subcaption}

\newcommand{\vtot} {\xi_{\textrm{tot}}} 
\newcommand{\vlin}{\xi_{\textrm{line}}}
\newcommand{\vhom}{\xi_{\textrm{hom}}}
\newcommand{\vel}{\nu_{\textrm{el}}}

\usepackage{amsmath,amssymb}

\usepackage[pdftex,colorlinks=true,bookmarks=false,citecolor=blue,urlcolor=blue]{hyperref} 

\begin{document}

\title{Inter-Core Crosstalk Impact of Classical Channels on CV-QKD in Multicore Fiber Transmission}
\vspace{-10pt}
\author{Tobias A. Eriksson, Benjamin J. Puttnam, Georg Rademacher, Ruben S. Lu\'is, \\ Masahiro Takeoka, Yoshinari Awaji, Masahide Sasaki, Naoya Wada}
\address{ National Institute of Information and Communications
	Technology (NICT), 4-2-1 Nukui-kitamachi, \\Koganei, Tokyo 184-8795, Japan.}\vspace{-3pt}
\email{eriksson@nict.go.jp}
\vspace{-15pt}

\begin{abstract}
Crosstalk-induced excess noise is experimentally characterized for continuous-variable quantum key distribution, spatially multiplexed with WDM PM-16QAM channels in a 19-core fiber. The measured noise-sources are used to estimate the secret key rates for different wavelength channels.
\end{abstract}
\vspace{-3pt}
\vspace{-20pt}

\section{Introduction}
\begin{figure}[b]
	\centering
	\vspace{-15pt}
	\includegraphics[width=0.95\textwidth]{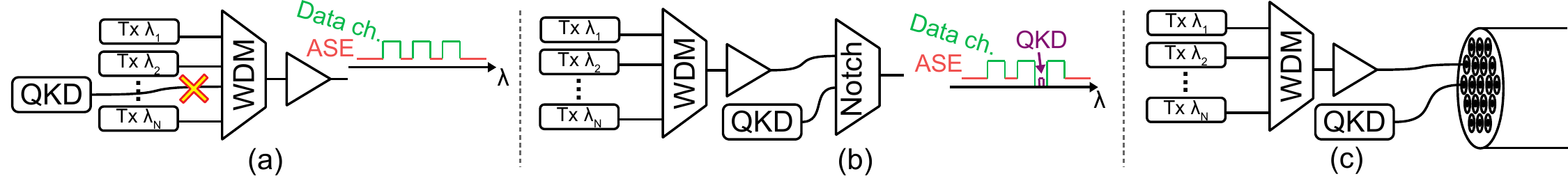}\vspace{-10pt}
	\caption{Illustration of multiplexing of QKD and classical channels: (a) EDFAs prohibits QKD channels to use a port of a WDM multiplexer. (b) QKD channels have to be multiplexed after the EDFA where the ASE must be suppressed for the QKD band. (c) System investigated in this paper, QKD channels use a dedicated core to avoid having to employ notch filtering.}
	\label{fig:exampleSmall}
	\vspace{-3pt}
\end{figure}

\vspace{-5pt}
Guaranteeing secure communication is one of the key challenges for next generation communication systems \cite{Cesare}. Quantum key distribution (QKD) is the only known technique for sharing a key between two remote parties with unconditional security \cite{Diamanti}. The generated keys can be used with either one-time pad encryption for unconditionally secure communication, or to replace public key agreements in conventional encryption schemes to improve the security. For widespread commercial use, QKD must likely be able to utilize the existing architecture to reduce the deployment cost. Co-propagation of QKD and classical channels over single mode fiber has been studied extensively \cite{Huang,Dynes,Mao,Karinou,ErikssonSUT,ErikssonXT}. However, current generation wavelength division multiplexing (WDM) systems typically employ erbium doped fiber amplifiers (EDFAs), generating amplified spontaneous emission (ASE) noise. A typical WDM transmitter structure is shown in Fig.~\ref{fig:exampleSmall}(a). Since the EDFA is placed after the WDM-multiplexer, the QKD signals cannot use one of the ports of the WDM multiplexer. A second multiplexing stage may be employed as shown in Fig.~\ref{fig:exampleSmall}(b) \cite{ErikssonXT}, where notch filtering of the ASE noise is applied before the QKD signals are combined with the classical signals. Multicore fibers (MCFs) are expected to be used in future systems and may open up possibilities to use the spatial domain for QKD \cite{DynesMCF,BaccoMCF}. For MCFs, notch filtering can be avoided by using one or more cores dedicated to QKD signals as indicated in Fig.~\ref{fig:exampleSmall}(c).

In this paper, we experimentally study the possibility of spatial multiplexing of CV-QKD signals and WDM multiplexed 24.5 Gbaud PM-16QAM signals in a 19-core MCF. We show that crosstalk from the classical channels prohibits secret key generation at the same wavelength. However, by assigning CV-QKD channels to wavelengths in the guard-bands between the classical channels, spatial multiplexing of CV-QKD and classical channels is possible. By estimating the CV-QKD performance based on our characterizations, we predict that spatial multiplexing with 17~Tbit/s (using 3-cores) classical WDM signals and 341 WDM-CV-QKD channels, assuming a conservative 5~GHz grid for 1~GHz CV-QKD channels in a single core.

\vspace{-8pt}
\section{Continuous Variable Quantum Key Distribution}\vspace{-5pt}
The secret key rate (SKR) of CV-QKD depends on assumptions made about the eavesdropper. Using reverse reconciliation and Gaussian modulated states, the SKR can be estimated as $SKR = \beta I_{AB} - I_{BE}$ under individual attacks, and  $SKR = \beta I_{AB} - \chi_{BE}$ under collective attacks \cite{Lodewyck}. $I_{AB}$ denotes the mutual information between Alice and Bob, $I_{BE}$ between Bob and Eve. They are given as
$I_{AB} = \frac{1}{2} \log_2 \left (\frac{V+\vtot}{1+\vtot} \right)$,  $I_{BE} = \frac{1}{2}\log_2 \left ( \frac{V_B}{V_{B|E}}\right ),$
where $V = V_{A} + 1$, $V_{A}$ is the modulated variance of the Gaussian distribution at Alice's side and $\vtot = \vlin + \vhom/T$ the total added noise between Alice and Bob, with $\vlin = 1/T-1-\epsilon$ as the total channel added noise. $T$ denotes the channel transmittance and $\epsilon$ the excess noise which in our experiments includes the noise introduced by crosstalk. Further, $\vhom = (1+\vel)/\eta-1$ is the total noise of the single quadrature receiver, where $\vel$ is the electrical noise added and $\eta$ the detection efficiency. Furthermore, $V_B = \eta T (V+\vtot)$ denotes the variance measured at Bob's receiver and $V_{B|E} = \eta ( \frac{1}{T(1/V + \vlin)} + \vhom)$ is Eve's conditional variance \cite{Lodewyck}. For the collective attacks, $\chi_{BE}$ denotes the Holevo information between Bob and Eve and is given as $\chi_{BE} = G(\frac{\lambda_1-1}{2}) + G(\frac{\lambda_2 -1}{2}) - G(\frac{\lambda_3 -1}{2}) - G(\frac{\lambda_4 -1}{2})$, where $G(x)\!=\!(x\!+\!1)\log_2(x\!+\!1)-x \log_2(x)$. Here, $\lambda_n$ denotes the symplectic eigenvalues and are given as $\lambda_{1,2}^2 = \frac{1}{2} (A \pm \sqrt{A^2 -4B})$, where $A = V^2(1-2T) + 2T + T^2(V+\vlin)^2$ and $B = T^2(V\vlin+1)^2$. Further,  $\lambda_{3,4}^2 = \frac{1}{2} (C \pm \sqrt{C^2 -4C})$, where $C=\frac { V\sqrt{B} + T(V+\vlin) +A \vhom} { T(V+\vtot) }$ and $D = \sqrt{B} \frac{ V + \sqrt{B} \vhom  } { T(V+\vtot)}$ \cite{Lodewyck}.

\vspace{-10pt}
\section{Experimental Setup and Measurement Procedure}
\begin{figure}[tb]
	  \centering
	  \includegraphics[width=\textwidth]{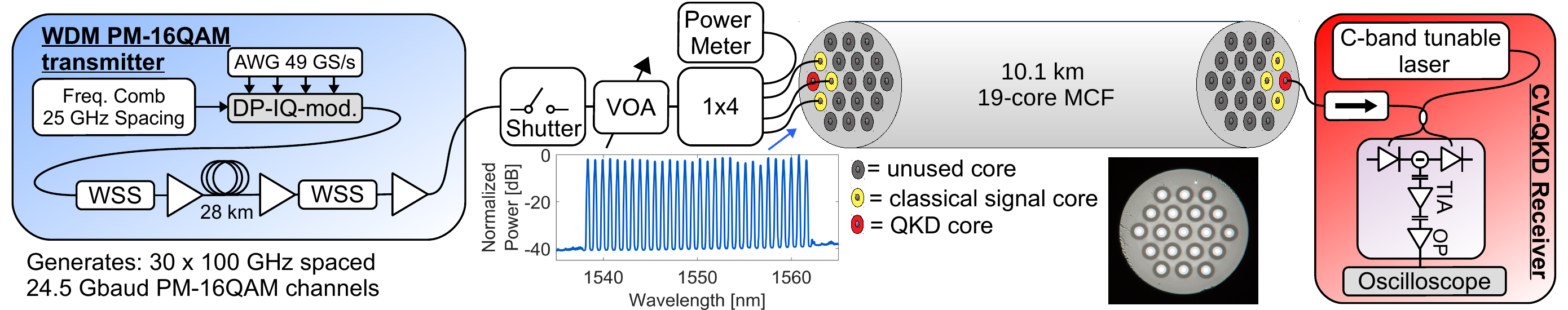}\vspace{-10pt}
	  \caption{Experimental setup showing: The transmitter generating 30 WDM channels of 24.5~Gbaud PM-16QAM with 100~GHz channel spacing. The multicore fiber with the QKD core surrounded by 3 cores loaded with the classical WDM channels. The QKD receiver based on a balanced photodetector with a C-band tunable laser as local oscillator. The insets shows the measured WDM spectra of the classical channels in one core, and the cross section of the MCF used in this experiment.  }
	  \label{fig:expsetup}
	  \vspace{-20pt}
\end{figure}

\vspace{-5pt}
The experimental setup is shown in Fig.~\ref{fig:expsetup}. The classical channels are generated with an optical frequency comb producing 25 GHz spaced carrier lines as light source, modulated with 24.5~Gbaud PM-16QAM signals using an arbitrary waveform generator (AWG) and a dual-polarization I/Q-modulator. Using wavelength selective switches (WSSs), channels are removed to create a channel spacing of 100~GHz. The spectrum of the transmitted signal is shown in the inset in Fig.~\ref{fig:expsetup}. The WDM signals are sent to a shutter followed by a 1-by-4 splitter where 3 outputs are decorrelated and launched in cores surrounding the CV-QKD test core in a 10.1~km 19-core MCF \cite{Sakaguchi}. The fourth output of the splitter is used to monitor the launched power. Since CV-QKD is extremely sensitive to impairments, we use a partition of the outer cores for CV-QKD signals, since they suffer less from crosstalk. In this experiment, we choose one of the six cores with the lowest numbers of neighboring cores for CV-QKD, as indicated in Fig.~\ref{fig:expsetup}. Such a configuration would allow for 6 cores for QKD and 13 cores for classical WDM coherent signals. The crosstalk has been shown to be dominated by the neighboring cores \cite{PuttnamXT}, which is why we can avoid loading all the cores to reduce the equipment demand.

The CV-QKD receiver is based on a balanced photodetector with low-noise electrical amplifiers and an analog bandwidth of approximately 1.6~GHz. The accumulated crosstalk in the CV-QKD core passes through an optical isolator before being mixed with the local oscillator (LO) from a C-band tunable laser source with 100~kHz nominal linewidth. The LO power on each photodiode is approximately 8 dBm but is found vary slightly with wavelength. The electrical signal is digitized with a real-time oscilloscope with 16~GHz bandwidth and 6.25~GS/s sampling rate. A 1~GHz low-pass filter is applied in the digital domain. Initially, the receiver electrical noise variance is determined in the absence of LO and crosstalk. Next, the shot noise variance of the receiver is evaluated after turning on the LO. In the final step, the excess noise from the crosstalk is measured by launching the classical channels in the adjacent cores. This procedure is repeated for each investigated wavelength.

\begin{figure}[b]
	\vspace{-18pt}
	\centering
	\begin{subfigure}{.348\textwidth}
		\includegraphics[width=\textwidth]{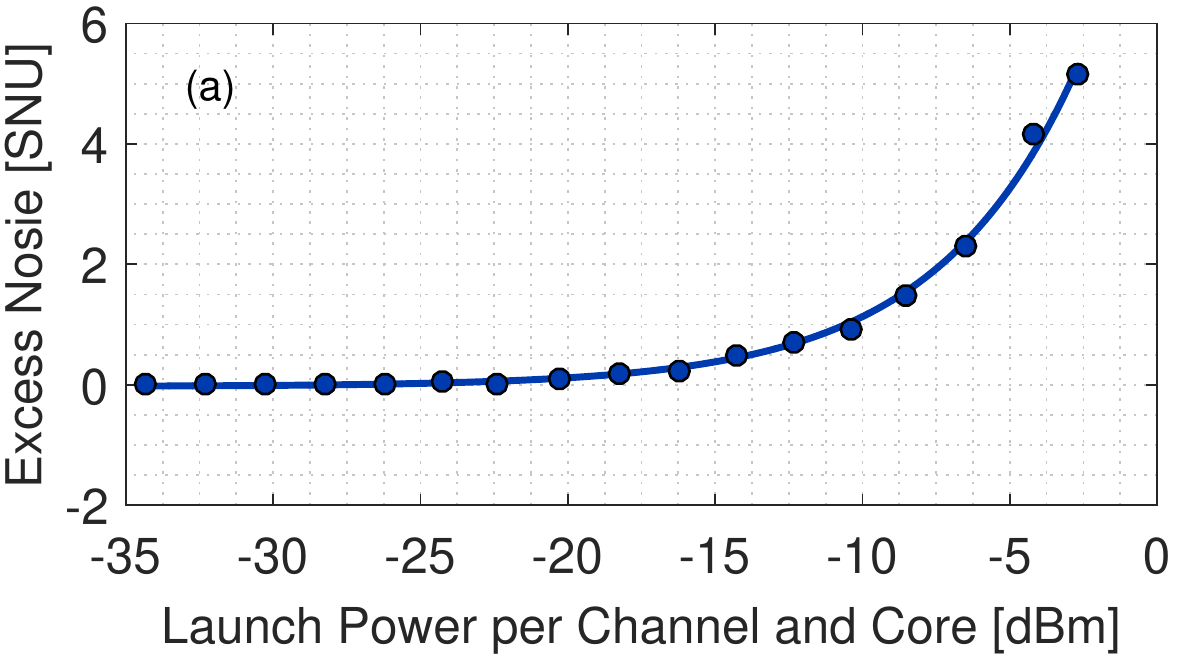}
	\end{subfigure}
	\begin{subfigure}{.622\textwidth}
		\includegraphics[width=\textwidth]{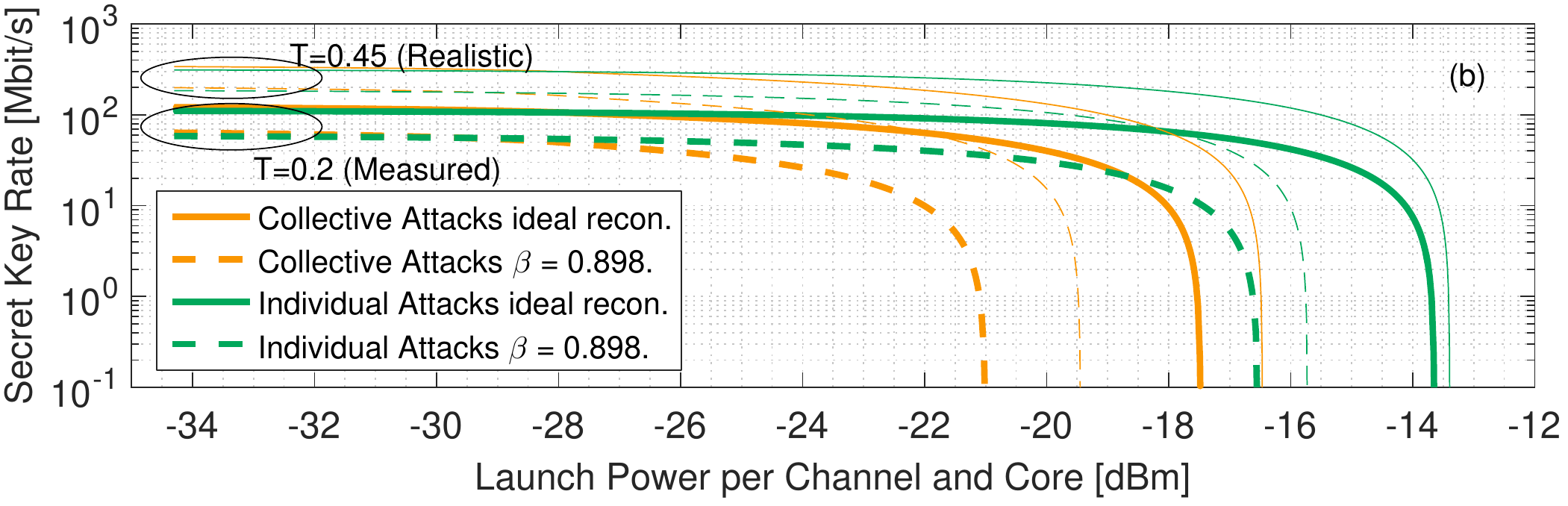}
	\end{subfigure}
	\vspace{-10pt}
	\caption{(a) Measured excess noise in shot noise units (SNU) for the CV-QKD channel at 1550.35~nm (fully overlapping with one PM-16QAM channel) as a function of launch power per channel and core. (b) Calculated secret key rates (SKRs) under collective and individual attacks, with ideal reconciliation and with a reconciliation efficiency of $\beta$ = 0.898. Two cases are assumed: The measured transmittance of T = 0.2 and an estimated  realistic transmittance of T = 0.45 (assuming 0.2~dB/km loss and that the coupling loss in the transmitter cannot benefit an eavesdropper.)}
	\label{Fig:powSweep} \vspace{-12pt}
\end{figure}

\vspace{-8pt}	
\section{Results}\vspace{-5pt}
We tune the LO to 1550.35~nm, which is the center of one of the WDM PM-16QAM channels. We vary the launched power of the classical signals and measure the excess noise from the crosstalk. The result is shown in Fig.~\ref{Fig:powSweep}(a). We use the measured excess noise to calculate the SKRs as a function of the launched power per channel and core. For these calculations we use and $\eta = 0.7$ and the measured receiver electrical noise of 0.08$N_0$ (at this specific wavelength) where $N_0$ is the measured shot noise variance. We use the measured transmittance of $T$ = 0.2, which includes coupling losses. Since the loss in our experiment is dominated by the coupling losses from non-ideal components, we also assume a more realistic loss of 0.2~dB/km and 1.5~dB total coupling losses. Here, the loss on the transmitter side can be considered inside Alice's transmitter and therefore only acts as an attenuation which can be taken into account when adjusting the transmitted power, resulting in $T$ = 0.45. (Note that we assume the same measured excess noise from crosstalk). The SKRs for $T$ = 0.2 and 0.45 are shown in Fig.~\ref*{Fig:powSweep}(b) for individual and collective attacks, with ideal reconciliation and with a reconciliation efficiency of $\beta$ = 0.898. The SKRs for all assumptions approach 0 for very low launch power per core, with the highest tolerable launch power being approximately $-$13~dBm for individual attacks with ideal reconciliation. These powers are 10 to 20~dB lower than typical values for 24.5~Gbaud PM-16QAM, indicating that due to crosstalk, transmitting CV-QKD signals and classical signals at the same wavelength in neighboring cores, is not feasible in this type of MCF.

\begin{figure}[tbp]
	\centering
	\begin{subfigure}{.99\textwidth}
		\includegraphics[width=\textwidth]{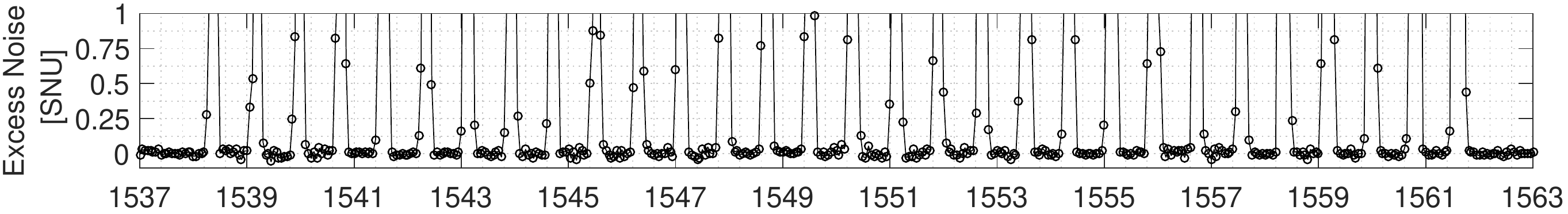}
	\end{subfigure}
	\begin{subfigure}{.99\textwidth}
		\hspace{1pt}
		\includegraphics[width=\textwidth]{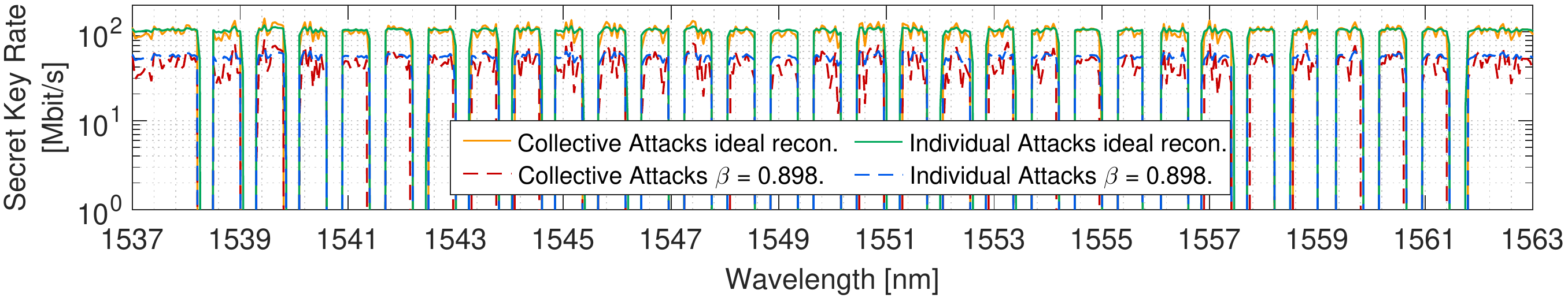}
	\end{subfigure}
	\vspace{-12pt}
	\caption{Top figure shows the measured excess noise in shot noise units (SNU) as a function of wavelength when the classical channels are launched with 12.2~dBm power per core. Bottom figure shows the corresponding estimated secret key rates under collective and individual attacks, with ideal reconciliation and with a reconciliation efficiency of $\beta$ = 0.898.}
	\label{Fig:wavelengtSweep}
	\vspace{-25pt}
\end{figure}

We also measure the excess noise at different wavelength by sweeping the receiver LO wavelength from 1537~nm to 1563~nm. The total launch power per core is 12.2~dBm for the 30 WDM channels of 24.5~Gbaud PM-16QAM on a 100~GHz grid, corresponding to $-$2.6~dBm per channel.  In Fig.~\ref{Fig:wavelengtSweep}, the measured excess noise is plotted as a function of wavelength. When the CV-QKD wavelength coincide with a classical channel the excess noise from crosstalk is drastically increased. The corresponding SKRs are also shown in Fig.~\ref{Fig:wavelengtSweep}. For the wavelengths where key generation is possible using the measured transmittance of $T$ = 0.2, the average SKR is 100~Mbit/s for collective attacks and 102~Mbit/s for individual attacks, both with ideal reconciliation. When assuming a reconciliation efficiency of 0.898, the corresponding SKRs are 46~Mbit/s and 52~Mbit/s for collective and individual attacks, respectively. 

If we assume WDM of CV-QKD channels on a 5~GHz grid, we can fit a minimum of 11 channels in the band between the classical channels. If we also assume placing 11 channels before and after the outermost classical channels, a total of 31 bands of CV-QKD channels can be transmitted. This sums up to to a SKR of 15.7~Gbit/s per core for the most conservative assumption of collective attacks and a reconciliation efficiency of $\beta$ = 0.898. If all six outer cores are used, the SKR may potentially be increased to 94.2~Gbit/s with a classical datarate of 70~Tbit/s, not including the center core since the impact from crosstalk from this core was not covered in our experiments. 

\vspace{-11pt}
\section{Conclusions}\vspace{-7pt}
We have measured the excess noise from the crosstalk of 100~GHz spaced WDM 24.5~Gbaud PM-16QAM signals on 1~GHz CV-QKD channels spatially multiplexed in a 19-core fiber. The CV-QKD signals can be placed at wavelengths in the guard-band between the classical channels and has the potential to support 341 QKD channels with 5~GHz spacing between 1537~nm and 1563~nm together with 17~Tbit/s classical data-rate in the three neighboring cores.
 
\vspace{3pt}
{\noindent  \footnotesize{ \hspace{-4.5pt} \emph{This work was partly funded by ImPACT Program of Council for Science,  Technology and Innovation (Cabinet Office, Government of Japan) and the Swedish Research Council (Vetenskapsr\aa det), Grant No 2017-06179.}\par}	\vspace{-9pt}
}

\end{document}